\documentclass[preprint,showpacs,preprintnumbers,amsmath,amssymb]{revtex4}

\usepackage{graphicx}% Include figure files
\usepackage{dcolumn}% Align table columns on decimal point
\usepackage{bm}% bold math
\usepackage{color}

\begin{document}

\title{Spatial Scaling in Model Plant Communities}

\author{Tommaso Zillio}
\affiliation{International School of Advanced Studies(SISSA), INFM,
via Beirut 2-4, 34014 Trieste, Italy}

\author{Igor Volkov}
\author{Jayanth R. Banavar}
\affiliation{Department of Physics, The Pennsylvania State
University, 104 Davey Laboratory, University~Park, PA 16802, USA}

\author{Stephen P. Hubbell}
\affiliation{Department of Plant Biology, The University of Georgia,
Athens, GA 30602 USA and The Smithsonian Tropical Research
Institute, Box 2072, Balboa, Panama}

\author{Amos Maritan}
%\email{maritan@sissa.it}
\affiliation{Dipartimento di Fisica `G. Galilei', Universit\`a di
Padova and INFM, via Marzolo 8, 35131 Padova, Italy}

\date{\today}

\begin{abstract}
We present an analytically tractable variant of the voter model that
provides a quantitatively accurate description of $\beta$-diversity
(two-point correlation function) in two tropical forests. The model
exhibits novel scaling behavior that leads to links between
ecological measures such as relative species abundance and the
species area relationship.
\end{abstract}


\maketitle

An ecological community represents a formidable many-body problem --
one has an interacting many body system with imperfectly known
interactions and a wide range of spatial and temporal scales. In
tropical forests across the globe, ecologists recently have been
able to measure certain quantities such as the distribution of
relative species abundance (RSA), the species area relationship
(SAR), and the probability that two individuals drawn randomly from
forests a specified distance apart belong to the same species (also
called $\beta$-diversity). In this letter, we take a first step
towards the development of an analytically tractable model that,
despite its simplicity, leads to a remarkably accurate quantitative
description of $\beta$-diversity in two different tropical forests.
It also indicates the existence of novel scaling behavior, revealing
previously unexpected relationships between $\beta$-diversity, RSA,
and SAR. The model we study is a version of the well-known voter
model \cite{Holley1} which has been applied to a variety of
situations in physics and ecology
\cite{Durrett1,Silvertown1,Hubbell1}.

Quite generally, one may study an ecosystem in $d$ spatial
dimensions with the most common case corresponding to $d=2$. We will
consider a hyper-cubic lattice in $d$ dimensions with each site
representing a single individual and where the lattice spacing,
$\sigma$, is such that $\sigma^d$  is of the order of the average
volume (area in $d=2$) occupied by an individual. At each time step
an individual chosen at random is killed and replaced, with
probability $1-\nu$ with an offspring of one of its nearest
neighbors or, with probability $\nu$, with an individual of a
species, not already present in the system. This last process is
called \emph{speciation} and the parameter $\nu$ is called the
\emph{speciation rate}. The case with $\nu=0$ is special (the
$\nu\rightarrow0$ case is distinct from $\nu\equiv0$) and has been
thoroughly studied (see, for example,
\cite{Frachebourg1,Liggett1,Liggett2,Durrett2}) and on a finite
lattice results in a stationary state with just a single species
(monodominance). The case $\nu=1$, in which a new species is
generated every time step, is trivial. Our focus is on $0<\nu<1$,
which is ecologically relevant and has been studied before
\cite{Durrett1,Chave3}, but is not well understood.

Let $F_{\vec x}^t$ be the probability that two randomly drawn
individuals separated from each other by $\vec{x}$ at time $t$ are
of the same species (for simplicity, the system is assumed to be
translationally invariant). The master equation for a community of
size $N$ occupying an area $A=\sigma^d N$ may be written as
\cite{rrr2}
\begin{equation}\label{f}
F_{\vec x}^{t+1} = \left(1-\frac{2}{N}\right)F_{\vec x}^t +
\frac{1-\nu}{N d} \sum_{i=1}^d \left(F_{\vec x+\sigma\vec e_i}^t +
F_{\vec x-\sigma\vec e_i}^t\right),
\end{equation}
where $\{\vec e_i\}$ is the basis vector set and $\vec x\neq0$. When
$\vec x=0$ $F_0^t = 1$.

In the continuum limit, Eq.(\ref{f}) becomes
\begin{equation} \label{masterFcont}
\frac{\partial F(r,t)}{\partial t} = \Delta_r F(r,t) - \gamma^2
F(r,t)+c\delta^d(r),
\end{equation}
where $r=|\vec{x}|$, $F(r,t) \equiv \sigma^{-d}F_{\vec x}^t$,
$\gamma^2 =2d\nu/\sigma^2$ and the time is measured in units of
$\tau = \sigma^2/(2dN)$. The continuum limit is obtained on choosing
$N\rightarrow\infty$, $\sigma\rightarrow0$ and $\nu\rightarrow0$ in
such a way that $\sigma^dN$ and $\gamma$ approach constant values
and $\tau\rightarrow0$. The first term on the right hand side
represents dispersal or diffusion. The second term is a decay term
arising from the effects of speciation \cite{Hubbell1}, whose
coefficient, $\gamma^2$, could generally be a function of $r$. The
last term is a consequence of the fact that for the discrete case,
Eq(\ref{f}), at $x=0$ one necessarily has the same species by
definition. The constant $c$ is fixed such that $\int_{C_0}
\mathrm{d}^dr \, F(r)=1$ where $C$ is the cube of side $\sigma$ (the
average nearest neighbor plant distance) centered in the origin
\cite{rrr3}.

\begin{table}
\begin{tabular}{|c|c|c|c|}
\hline $d$  &  $a$  &  $b$  &  $z'$   \\
\hline
$1$  & $0.5$ & $1.05$ & $0.03$ \\
$2$  & $0.87$ & $1.2$ & $0.3$  \\
$3$  & $0.95$ & $1.21$ & $0.5$ \\
\hline
\end{tabular}
\caption{Scaling exponents for $d=1,2,3$ determined from the scaling
collapse of the normalized RSA and SAR plots (Figs. 1 and 2).
\label{exponents}}
\end{table}

The stationary solution of Eq.(\ref{masterFcont}) is
\begin{equation}\label{continuum}
F(r) = s \,r^{\frac{2-d}{2}}K_{\frac{2-d}{2}}(\gamma r),
\end{equation}
where $K$ is the modified Bessel function of the second kind
\cite{Abramowitz1}; $s=c(2\pi)^{-d/2}\gamma^{\frac{d-2}{2}}$.
We have carried out extensive simulations on a square lattice and on a
hexagonal lattice with periodic boundary conditions in both cases
and have verified that the results are in excellent accord with the
analytic solution and independent of the microscopic lattice used.

When $d=1$ the above expression takes a simple form: $F(r) \sim
\exp(-\gamma r)$. It is easy to verify that the stationary solution
of the discrete equation (\ref{f}) with $d=1$ is also an exponential
function $F_{\vec x} = e^{-|\vec x|/\lambda}$, where $\lambda =
\sigma\left[ \ln \frac{1-\nu}{1-\sqrt{\nu(2-\nu)}} \right]^{-1}$ is
the correlation length. In the small $\nu $ limit $\lambda \sim
\gamma^{-1}\sim 1/\sqrt{\nu}$ and one may identify a natural scaling
variable $A\nu^{d/2}$.

A second scaling variable is identified by noting that
\begin{equation} \label{moments} \nonumber
\sum_{\vec x}F_{\vec x}=\frac{1}{N}\sum_{\vec x,\vec y}\left\langle
I_{\vec x,\vec y}\right\rangle=\frac{1}{N}\left\langle\sum_{i=1}^S
n_i^2\right\rangle=\frac{\langle n^2 \rangle}{\langle n \rangle } ,
\end{equation}
provides a characteristic scale for $n$, where the indicator
$I_{\vec x,\vec y}$ is a random variable which takes the value $1$
with probability $F_{\vec x- \vec y}$ and $0$ with probability
$(1-F_{\vec x- \vec y})$, $S$ and $n_i$ represent the number of
species and the number of individuals of the $i$-th species
respectively and $\langle\ldots\rangle$ is an ensemble average, or a
time average over a long time period. Furthermore, using
Eq.(\ref{f}) and its Fourier Transform, one finds that $\sum_{\vec
x}F_{\vec x}\approx\int F(r){\mathrm d}^d r\sim \nu^{-a}$ with
$a=\min(1,d/2)$. This results in a second natural scaling variable
$n \nu^a$.

Within the context of the same model, we turn to an investigation of
other quantities of ecological interest, notably the relative
species abundance (RSA) and the species-area relationship (SAR)
\cite{Hartefootnote,Harte2} in the limit of small speciation rate,
$\nu$, and large volume (area for $d=2$) $A$ of the system. Here we
will investigate the case when $\nu$ is small but the system is
still far from the onset of monodominance ($ A\nu^{d/2}>1$).

First, let us postulate a scaling form for the total number of
species $S(\nu,A)$ within an area $A$:
\begin{equation} \label{scalingS}
S(\nu,A) = A^{z'} \hat S(A\nu^{d/2}).
\end{equation}
The exponent $z'$ is equal to the traditional species-area
relationship exponent, $z$, only if $\hat S$ is constant.
From our numerical simulation we find that this is true when $A\ll\nu^{-d/2}$,
while when $A\gg\nu^{-d/2}$ $S(\nu,A)$ is linear in $A$.

\begin{figure}
\includegraphics[width = 0.8\textwidth]{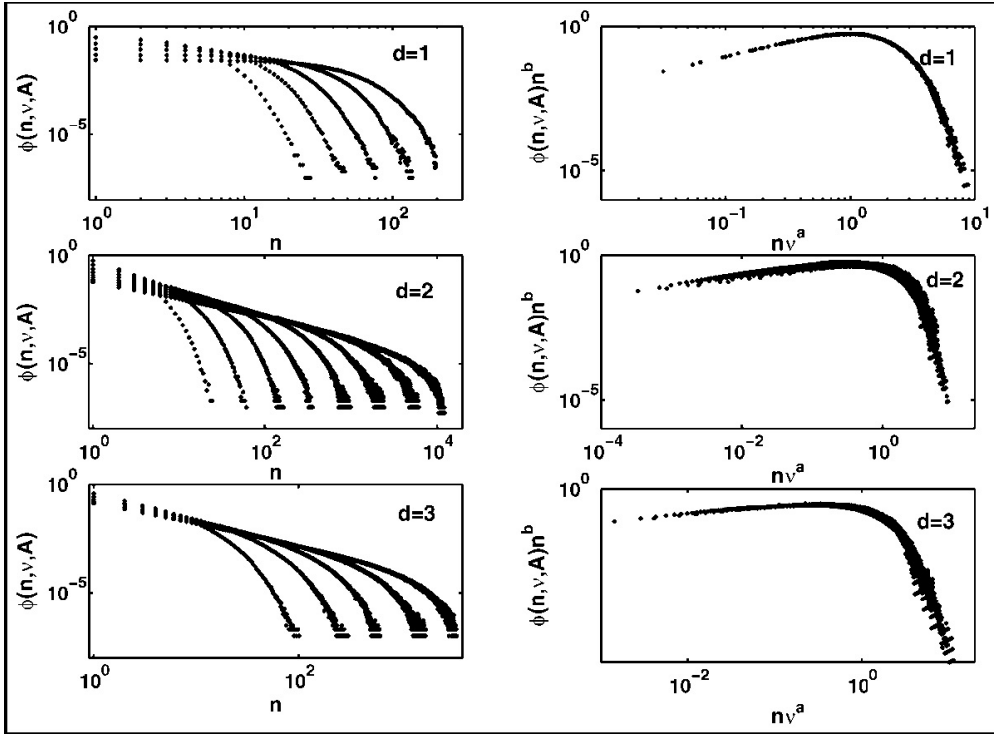}
\caption{\label{fig:rsa} Left column: plots of the normalized RSA
for $d=1,2,3$ with $\nu=0.001, 0.003, 0.01, 0.03, 0.1$ (the $d=2$
plot also shows the results for $\nu=0.0001, 0.0003, 0.3$);
{$L=200$.} Right column: plots of the data collapse yielding a
measure of the exponents $a$ and $b$ in Table 1. }
\end{figure}

\begin{figure}
\includegraphics[width = 0.8\textwidth]{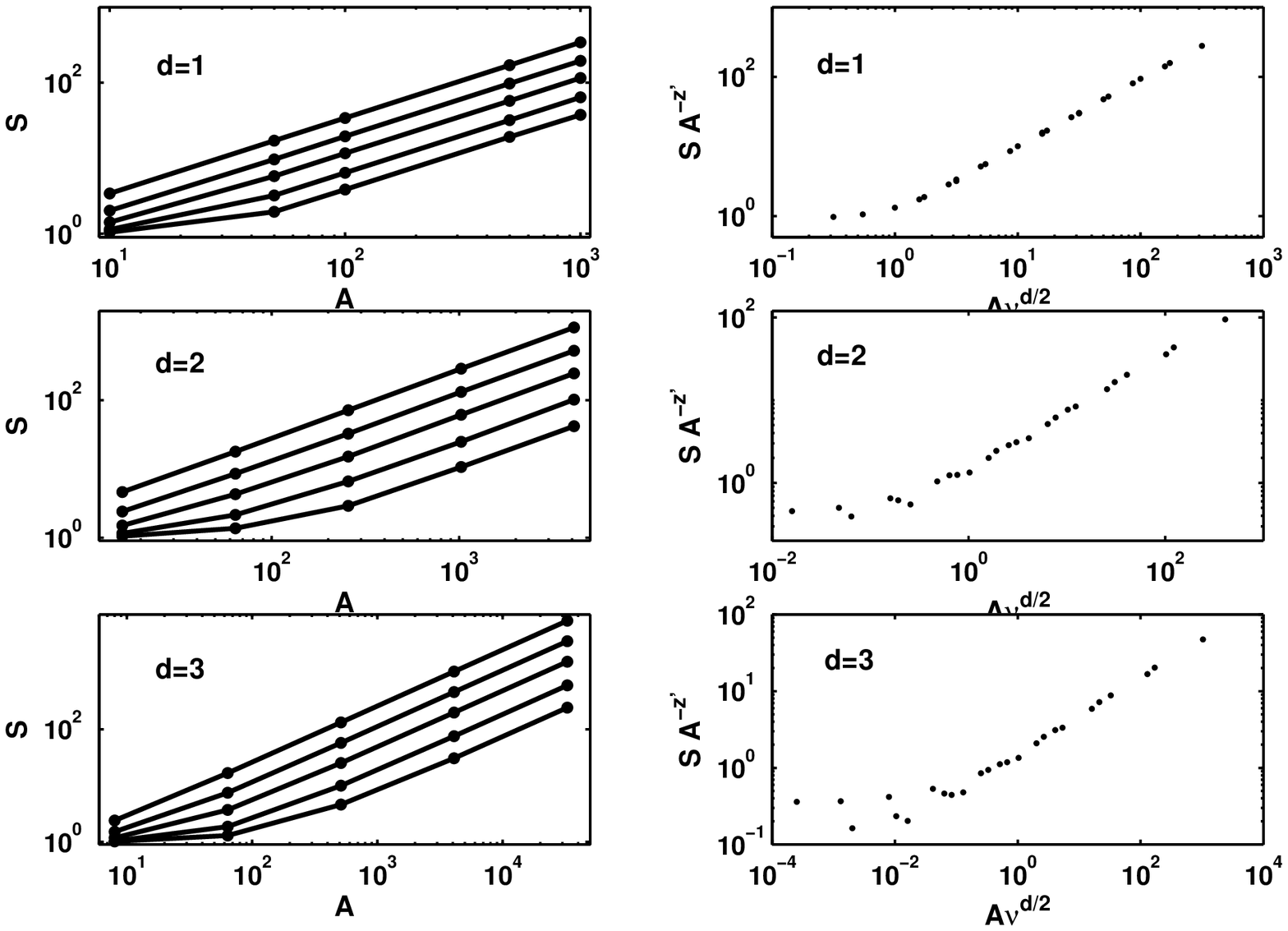}
\caption{\label{fig:sar} Left column: plots of the SAR for $d=1,2,3$
with $\nu=0.001, 0.003, 0.01, 0.03, 0.1$. Right column: plots of the
data collapse yielding a measure of the exponent $z'$ in Table 1.}
\end{figure}

Let us introduce a scaling form of the normalized RSA distribution
of species, $\phi(n,\nu,A) \equiv f(n,\nu,A)/S(\nu,A)$, where
$f(n,\nu,A)$ is the number of species with $n$ individuals:
\begin{equation} \label{scalingphi}
\phi(n,\nu,A) = n^{-b}\hat\phi(n\nu^a, A\nu^{d/2}),
\end{equation}
where $b$ is an as yet undetermined exponent. By definition,
\begin{eqnarray} \label{sphi}
\sum_n\phi(n,\nu,A)\equiv 1
\end{eqnarray}
and
\begin{eqnarray} \label{nphi}
\sum_n n\phi(n,\nu,A)\equiv A/S(\nu,A)
\end{eqnarray}
(for simplicity, we set $\sigma=1$ so that $A=N$).

\begin{figure}[h!]
\includegraphics[width = 0.5\textwidth]{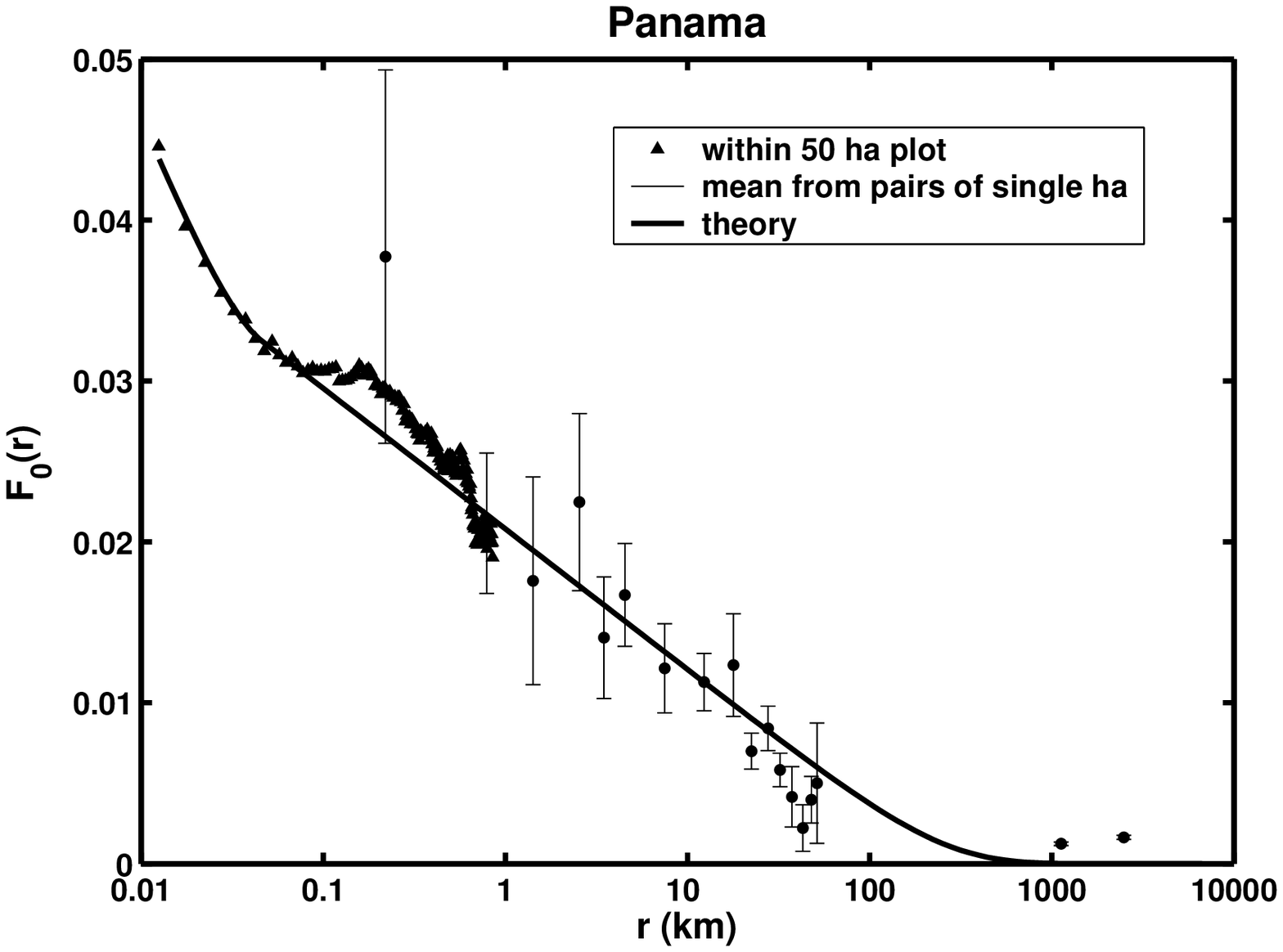}
\includegraphics[width = 0.5\textwidth]{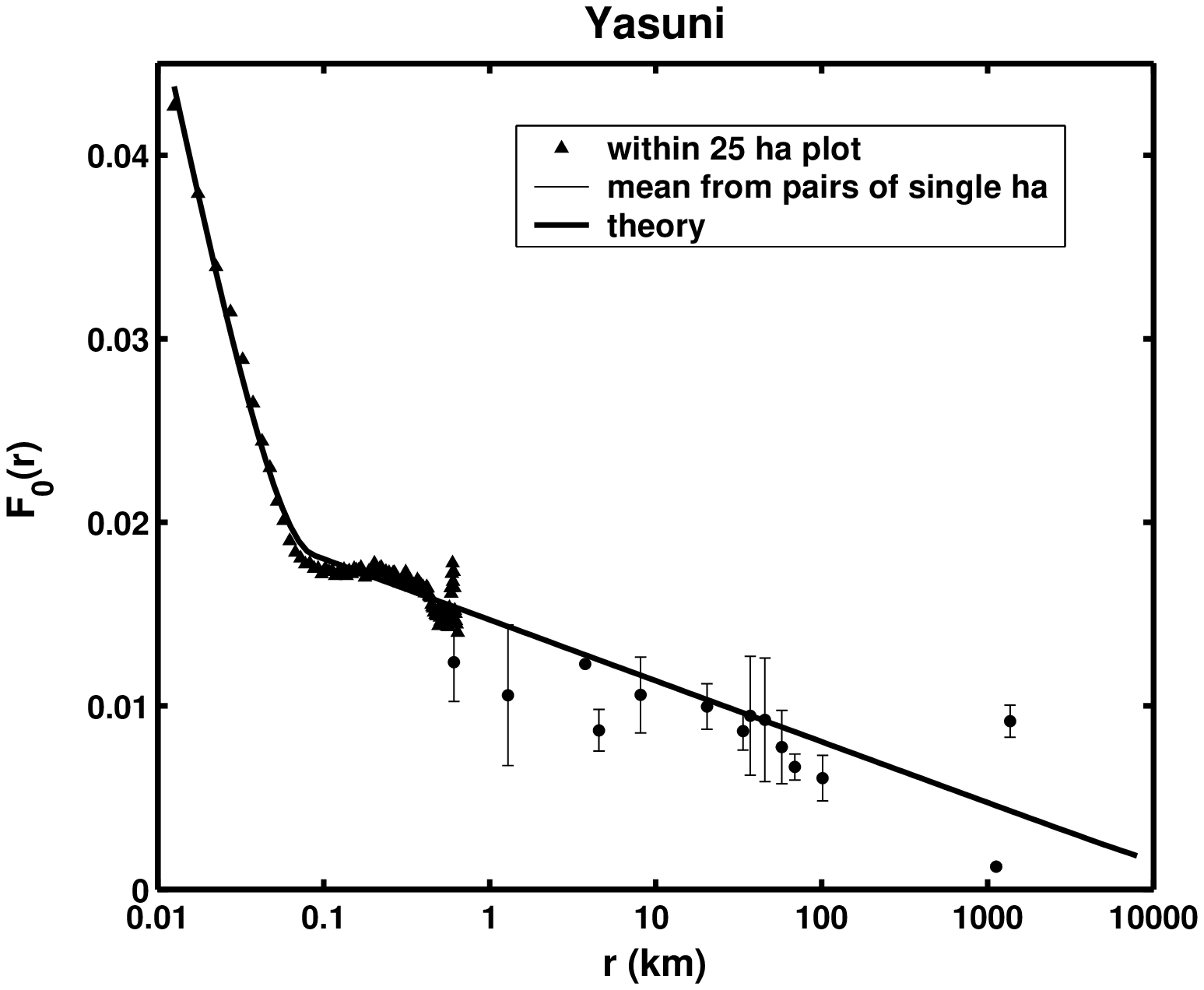}
\caption{\label{fig:beta} Beta Diversity data \cite{Condit1} along
with the best fits using Eq.(\ref{eq:beta}) for plots in  (a) Panama
($R=46$m, $\gamma_0^{-1}=68$m, $\gamma_1^{-1}=210$km and
$c_0=0.012$) and in (b) Ecuador-Peru (Yasuni) ($R=86$m,
$\gamma_0^{-1}=69$m, $\gamma_1^{-1}=23,500$km and $c_0=19$). The
Janzen-Connell effect pushes conspecific individuals further away
from each other and thus the probability function $F_0$ declines
more steeply within the zone of its operation than at larger
distances as in the data. Note that for length scales much bigger
than $R$, in a regime in which Janzen-Connell effects do not play a
significant role, the beta diversity curve decays much slower with
length in Yasuni than in Panama by nearly a factor of $100$. This
difference may be attributed to a very strong climatic gradient (in
annual rainfall) across the isthmus in Panama, which causes rapid
spatial changes in forest composition. In Yasuni, on the other hand,
there are large areas with very similar climate. The experimental
protocol and a description of the symbols are in \cite{Condit1}}
\end{figure}

The normalization condition Eq.(\ref{sphi}) becomes
\begin{equation}
1=\sum_{n\geqslant1}\phi(n,\nu,A)\sim
\nu^{a(b-1)}\int_{\nu^a}^\infty \mathrm dx
x^{-b}\hat\phi(x,A\nu^{d/2}).
\end{equation}
Assuming that $\hat\phi(x,y)$ does not diverge as $x$ tends to zero
(which is justified {\em a posteriori} by numerical simulations),
one finds that $b>1$ if $\hat \phi(x,y)$ approaches a constant value
(modulo multiplicative logarithmic corrections) and $b=1$ if $\hat
\phi(x,y) \rightarrow0$ as $x\rightarrow0$. In both cases
$\hat\phi(0,y)$ is independent of $y$ (also in accord with the
results of computer simulations). The condition on the average
population per species, Eq.(\ref{nphi}), leads to
\begin{eqnarray} \label{eq9}
\nu^{-d(1-z')/2}\hat{S}'(A\nu^{d/2}) \equiv
\sum_{n\geqslant1}n\phi(n,\nu,A) \\ \nonumber\sim
\nu^{a(b-2)}\int_{\nu^a}^\infty \mathrm dx
x^{1-b}\hat\phi(x,A\nu^{d/2}) ,
\end{eqnarray}
where $\hat S'(y)=y^{1-z'}/\hat S(y)$. Detailed simulations followed
by a scaling collapse indicate that in all dimensions, $b < 2$. It
then follows that the lower integration limit in the above integral
can be safely put to zero (a non-zero lower limit of integration
would merely result in corrections to scaling) leading to the
scaling relation
\begin{equation} \label{sc}
a(2-b) =  \frac{d}{2}(1-z')
\end{equation}
and
\begin{equation} \label{nphi2}
\hat S'(A\nu^{d/2}) = \int_0^\infty \mathrm dx
x^{1-b}\hat\phi(x,A\nu^{d/2}).
\end{equation}
The linear dependence of $S(\nu,A)$ on $A$ arises from $\phi$
becoming independent of $A$ in the limit $A\gg\nu^{-d/2}$ (see
Eq.(\ref{nphi})). This then leads to the scaling function $\hat S(y)
\sim y^{1-z'}$ in Eq.(\ref{nphi}). This also follows from noting
that $\hat S'$ approaches a constant value for large $A$
(Eq.(\ref{eq9})).

We have carried out extensive simulations with hypercubic lattices
of various sizes in $d=1,2,3$. A series of simulations with fixed
size and varying speciation rate was used for the determination of
the normalized RSA ($L=200$ for $d=2$ and $L=100$ for $d=1,3$, $L$
being the side of the hypercube used). Another series of
simulations, varying both the speciation rate and $L$ was carried
out to determine the SAR curves. $S(\nu,A)$ is the mean number of
species in a simulation with speciation rate $\nu$ on a hypercubic
lattice of size $A=L^d$. In order to carry out the collapse, we used
the automated procedure described in \cite{Bhattacharjee1}. Applying
this procedure to the data on the normalized RSA obtained in our
simulations, we were able to obtain the values of the exponents $a$
and $b$ (see Table \ref{exponents}, and Fig.\ref{fig:rsa}).

As expected, our extensive computer simulations indicate that $\phi$
is only weakly dependent on $A\nu^{d/2}$. In all dimensions the
scaling exponents (see Table \ref{exponents}) approximately satisfy
Eq.(\ref{sc}) and $b$ is in the interval $(1,2)$. Figure
\ref{fig:sar} shows a collapse plot which confirms the scaling
postulates above. The biggest deviation from our theory is found in
the value of $a$ in $d=2$, the upper critical dimension for
diffusive processes and are suggestive of logarithmic corrections.
Interestingly, our scaling relation seems to hold even for this
case.

We conclude with a generalization of our model which provides an
excellent quantitative fit to the $\beta$-diversity data of two
tropical forests. The key idea is that the factor $\gamma$ in
Eq.(\ref{masterFcont}) can, in principle, take on two distinct
values ($\gamma=\gamma_0$, $r<R$; $\gamma=\gamma_1$, $r\geqslant R$,
where $R$ is a characteristic length scale separating the two
distinct regimes). Such an effect arises physically from now
well-established density-dependent processes first hypothesized by
Janzen \cite{Janzen1} and Connell \cite{Connell1} in which the
survival probability for an offspring is decreased in the vicinity
of adults of the same species. Janzen and Connell postulated that
this increased mortality rate of seeds and seedlings near adults
arise from the presence of pests that are host-specific, i.e,
specialized to that type of tree, and experimental evidence supports
this conclusion \cite{Harms1}. The microscopic model that we
consider is slightly modified from the previous version in that
there is a non-zero probability that a new-born is immediately
killed with a probability proportional to the number of conspecific
adults within a circle of radius $R$ centered at the site.

The solution of Eq.(\ref{masterFcont}) in $d=2$ with two distinct
values of $\gamma$ is
\begin{equation}
F_0(r) = \left\{
\begin{array}{cl}
c_0 K_0(\gamma_0 r) + c_1 I_0(\gamma_0 r), & r<R \\ \\c_2 K_0
(\gamma_1 r), &  r>R
\end{array}
\right. \label{eq:beta}
\end{equation}
where $I_0$ and $K_0$ are  modified Bessel functions
\cite{Abramowitz1} and the constants $c_1$ and $c_2$ are fixed by
imposing the continuity of $F_0(r)$ and its derivative at $r=R$
($I_0(r)$ diverges as $r\rightarrow\infty$ and is therefore excluded
in the region $r>R$). Figure \ref{fig:beta} shows that our theory
leads to good fits of the data on $\beta$-diversity for tree
communities in both central Panama (top panel) and Ecuador-Peru
\cite{Condit1}.

\begin{acknowledgments}
This work was supported by COFIN MURST 2001, NASA, NSF Grant No.
DEB-0346488 and NSF IGERT grant DGE-9987589.
\end{acknowledgments}

\bibliography{ecology}% Produces the bibliography via BibTeX.

\begin{thebibliography}{19}
\bibitem[{rrr({\natexlab{b}})}]{rrr3}
\emph{\bibinfo{title}{{ In general $F_{\vec x}^t = \int_{C_x} \mathrm{d}^dr \,
  F(r,t)$ where $C_x$ is a cube of side $\sigma$ centered at $\vec x$. It is
  only when $\vec x \neq 0$ that $F_{\vec x}^t = \sigma ^d F(r,t)$ in the
  continuum limit. When $d \geqslant 2$, $F(r)$ diverges in the origin but
  $\int_{C_0} \mathrm{d}^dr \, F(r) $ is finite.}}}
\end{thebibliography}

\end{document}